\documentclass[10pt]{article}
\usepackage{amssymb,amsmath} 
\usepackage{graphicx}
\usepackage[colorlinks=true, pdfstartview=FitV, linkcolor=blue, citecolor=blue, urlcolor=blue]{hyperref}
\usepackage{bbm}
\usepackage{url}

\def\pl{\partial}

\newcommand{\bib}{\bibitem}

\newcommand{\ci}{\cite}

\newcommand{\harr}[1]{\smash{\mathop{\hbox to .5in{\ \rightarrowfill\ }}
      \limits^{#1}}}

\def\f{\phi}

\newcommand{\db}{{\,{\rm d}\kern-1.6ex-}}
\newcommand{\dir}{{\pl\kern-1.2ex {/}}}

\newcommand{\lra}{\leftrightarrow}

\newcommand{\plra}{\pl^{\kern-1.25ex^\lra}}

\def\XXint#1#2#3{{\setbox0=\hbox{$#1{#2#3}{\int}$}
     \vcenter{\hbox{$#2#3$}}\kern-.5\wd0}}

\def\bib#1{\bibitem[#1]{#1}}

\begin{document}

\title{Does a phase shift occur in an AC arc?
}

\author{{\Large Chas.~Proteus Steinmetz}\\ \\
\normalsize Translated by Gerald Kaiser\\
\normalsize \href{http://wavelets.com}{Center for Signals and Waves}\\ 
\normalsize kaiser@wavelets.com
}

\maketitle

\begin{abstract}\noindent
This is a translation of a classic paper \ci{S1892} in German showing that the \it apparent power \rm in an AC arc is larger than the \it active power \rm although no phase shift exists between the voltage and the current, indicating that the \it reactive power \rm vanishes. The phenomenon studied in this paper gave rise to a variety of mutually conflicting `power triangle' models relating the active, reactive, and apparent powers P, Q, and S whose merits are still under debate today.
\end{abstract}

In Issue 34 of the ETZ, J. Heubach derives the existence of a phase shift in an AC arc from several sets of observations, in which he simultaneously measures the voltage and current and -- using an electric dynamometer -- the consumed energy, and then computes as a cosine of the phase shift angle the quotient 
\begin{align*}
\frac{\rm Watt}{\rm Volt\times \text{Amp\`ere}}.
\end{align*}
It is striking that the ``phase shift angle'' so defined, within the observational limits, does not appear to depend on the period of the waveforms, as is generally the case for phase shifts, and the inference is obvious that the difference between Volt-Amp\`ere and Watt is not due to a phase shift at all. Indeed, the above determination method is valid only under the assumption that both the current intensity and the EMF are sinusoidal waves, a condition which cannot be fulfilled in the AC arc. For since the effective resistance in the arc depends on the magnitude of the current, it must vary periodically with twice the frequency of the AC source. An analytical study shows that, when an alternating current flows through a resistance varying at twice its frequency, a wave with triple the frequency is added to the simple sine wave of the EMF, or the current, or both; hence the current and the EMF cannot both be sinusoidal waves and the above method is therefore incorrect.

Very detailed studies of the processes in the AC arc, by the simultaneous determination of the current and voltage curves, were already made two years ago at Cornell University, Ithaca (New York State), together with a number of other features of the AC waveforms whose results for the understanding of alternating current phenomena are of the highest importance, so it is very striking that these investigations (published in the ``Transactions of the American Institute of Electrical Engineers'') are apparently little known in Germany.

Figure 11 represents the voltage, current and energy of an AC arc with 75 full waves per second, as measured by WB Tobey and GH Walbridge.\footnote{Transactions of American Institute of Electrical Engineers, 1890, vol. VII, No. 11: ÒInvestigation of the Stanley Alternative Current Arc Dynamo.Ó} 
The current came from a Westinghouse Arc light alternator, as has been reported at the time in the ETZ. It gives a self-regulating, almost perfectly constant current from short-circuit up to 3000V. To achieve self-regulation, this machine is designed to possess a very high self-inductance of the armature (American Institute of Electrical Engineers, Transactions November and December 1890). Due to the use of induction-free resistors, as well as induction coils, the voltage and current were very close to sine waves. The arc lamp produced the waves shown in the figure, where the current is slightly sharpened by the triple-periodic wave $ b\sin 3\f$ but has virtually retained the sine shape as a result of the high armature inductance, while the EMF has a saddle-shaped depression between two high peaks. A noticeable phase shift did not exist. 
\begin{figure}[h]
\begin{center}
\includegraphics[width=4.8 in]{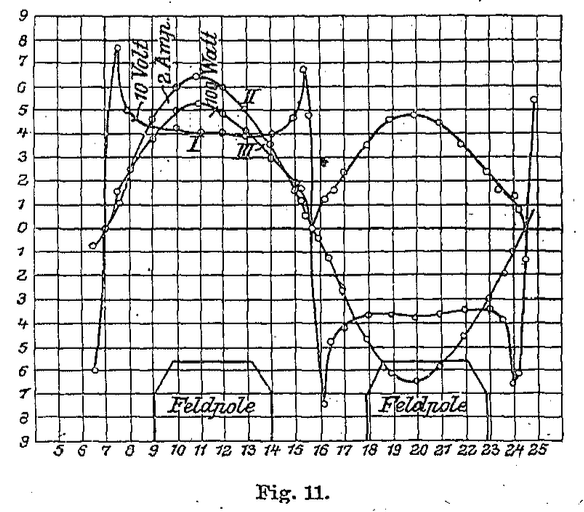}
\end{center}
\end{figure}

Conversely, a machine with infinitesimal armature reactance would keep the emf sinusoidal and give the current  a sharp tip at its center. 

The remarkable fact about the above curves is that the RMS current = 8.86A, RMS EMF = 42V, and instantaneous power (product of the instantaneous current and EMF) = 314Watt, whence the quotient
\begin{align*}
\frac{\rm Watt}{\rm Volt\times \text{Amp\`ere}}=\frac{314}{8.86\times 42}=0.845
\end{align*}
gives an apparent phase shift of 32$\,^{\circ}$. Nevertheless, no phase shift exists.

Thus:

The calculation of a phase shift by using an electrical dynamometer is only accurate for sine waves.

In arcs, the current and voltage cannot both be sine waves, and the electric dynamometer method is therefore inapplicable. 

Investigations of the electric arc show no appreciable phase shift between current and voltage, but they do show a significant difference between Volt-Amper\`es and Watts, and as a result of this difference, one of the two waveforms must strongly deviate from a sine wave; this deviation is due to the arc.

We further note that Dr. O. Fr\"ohlich has demonstrated by direct observation using his well-known method that a phase shift does not occur in an AC arc.

\section*{Acknowledgements}
I thank Dimitri Jeltsema for bringing this important paper to my attention and Hans van den Berg, David Griffiths, Thorkild Hansen, Friedrich Hehl, and Michael Kiessling for help in its translation.
--- Gerald Kaiser

\end{document}